# Attribute Compression of 3D Point Clouds Using Laplacian Sparsity Optimized Graph Transform

Yiting Shao[1], Zhaobin Zhang[2], Zhu Li[3], Kui Fan[4], Ge Li[5]
School of Electronic and Computer Engineering, Peking University Shenzhen Graduate School
[1]ytshao@pku.edu.cn, [4]kuifan@pku.edu.cn, [5]gli@pkusz.edu.cn
Computer Science & Electrical Engineering Department, University of Missouri – Kansas City,
[2]zzktb@mail.umkc.edu, [3]zhu.li@ieee.org

*Abstract*— **3D sensing and content capture have made significant progress in recent years and the MPEG standardization organization is launching a new project on immersive media with point cloud compression (PCC) as one key corner stone. In this work, we introduce a new binary tree based point cloud content partition and explore the graph signal processing tools, especially the graph transform with optimized Laplacian sparsity, to achieve better energy compaction and compression efficiency. The resulting rate-distortion operating points are convex-hull optimized over the existing Lagrangian solutions. Simulation results with the latest high quality point cloud content captured from the MPEG PCC demonstrated the transform efficiency and rate-distortion (R-D) optimal potential of the proposed solutions.**

## I. INTRODUCTION

With the rapid development of 3D data acquisition technologies, point clouds are becoming an effective way to express the surfaces of 3D objects and scenes [1]. Compared with traditional 2D images and videos, point clouds are usually unorganized distributed in 3D space without structured grids, and different point cloud frames may have different number of points. Considering the huge amount of data and band-limited networks, point cloud compression has been becoming a critical and challenging research topic.

There have been some work on point cloud compression. Motivated by 3D mesh coding in [2], [3] applied octree structure for point cloud compression. Later, [4] presented a generic scheme with octree for progressive point cloud coding. [1] extended octree for dynamic point cloud compression. Octree partition is an effective way for point cloud geometry coding, however, for attribute compression, it cannot exploit the correlation among points well [5].

To solve this problem, Zhang *et al.* in [6] constructed graphs at a certain level of octree and use graph transform to encode point cloud attributes. The transform scheme had better performance over traditional DCT that reported significant improvement in point cloud compression. The way to construct the graph would create many isolated sub-graphs when point cloud is sparse. To tackle the problem, [7] used K-nearest-neighbor (KNN) method to connect more distant points in a graph. However, the KNN graph is not guaranteed to construct all points of a block in one graph, thus it maybe not an efficient way to reduce sub-graphs. Ricardo *et al.* proposed the region-adaptive hierarchical transform (RAHT) for attribute compression, but graph transform outperforms RAHT in many tests from results in [8]. Rufael *et al.* mapped the color attributes to a JPEG grid and used the existing JPEG codec for color compression [5]. It would be more computationally efficient than graph transform, but may not be better on compression performance. Graph transform is current the state-of-the art on compression performance for attribute coding, but it still exists some issues, such as the sub-graphs.

In this paper, we propose an optimized scheme of graph transform on point cloud attribute compression. It is assumed that point clouds have been geometry compressed based on octree. K-dimension (k-d) tree partition is applied to split all points evenly into transform blocks. We connect all points to form a graph in each transform block and the edge weights are optimized by two trained parameters, which influence the Laplacian sparsity on the adjacency matrix for graph transform. A Lagrangian rate-distortion optimization (RDO) is utilized to specify the quantization mode. Experimental results demonstrate that our method has better transform efficiency and R-D performance than traditional DCT based method.

The rest of this paper is organized as follows. The details of the proposed method are explained in Section II. In Section III, we present some experimental results to evaluate the proposed scheme. Finally, we conclude in Section IV.

## II. PROPOSED METHOD

The reference software solution of the MPEG PCC adhoc group [9] is based on octree for geometry compression and the coordinates of points can be reconstructed at the decoder. Inspired by this scheme, we assume the geometry is coded via separate pipeline and we code attributes as graph signals on the reconstructed geometry. The optimization mechanism is mainly embodied in the partition of transform blocks, Laplacian sparsity optimization for graph transform and the selection of quantizing mode.

### A. Point Cloud Partition via K-d Tree

K-d tree is a binary data-partitioning tree for organizing points in a *k*-dimensional space [10]. It represents a hierarchical subdivision of space using splitting hyperplanes that are perpendicular to the corresponding axes. While building the k-d tree, choice of the dimensions to split and the splitting points are two major factors affecting the data structure [11]. For the choice of dimensions, one method chooses the dimension in a round-robin fashion and another chooses the widest spread dimension. About the selection of points as splitting nodes, one method uses the midpoint of the dimension width as the splitting point and another chooses the median point. The former divides the dimension into two parts with equal width while the latter makes the number of points almost equal in two parts.

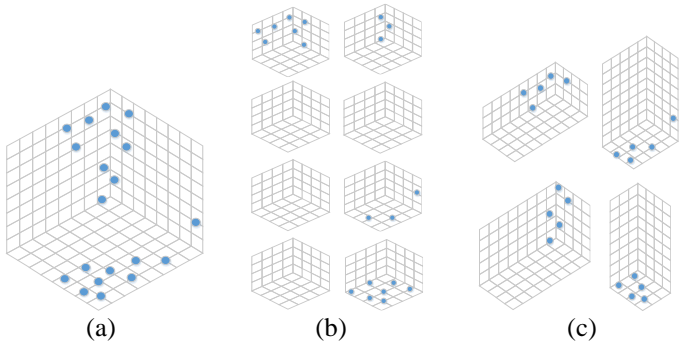
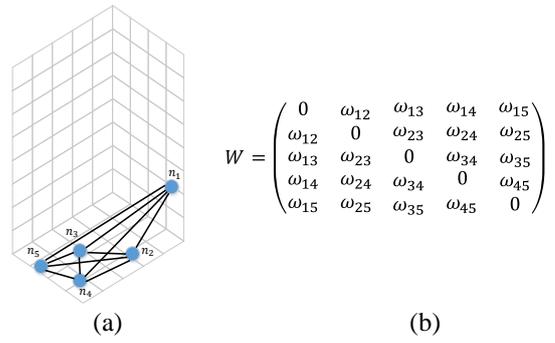

Fig. 1: (a) Example of a sparse point cloud. (b) Octree partition. (c) K-d tree partition.

Fig. 2: (a) Example of a graph in a transform block. (b) An adjacency matrix for the graph.

Octree is another common space decomposition tree in point clouds. For attributes compression, some previous work applied octree to get transform blocks and construct a graph for each block [6] [7]. The number of points in each block is almost different, even to be zero. It would result in too many isolated sub-graphs if the point cloud is sparse. Therefore, we develop the median-based k-d tree to divide points along the widest spread dimensions for blocks partition. The widest spread dimension represents that points in this dimension are of weaker correlation and the median-based partition makes the number of points in each k-d tree nodes almost the same. Comparison of octree and k-d tree partition for a sparse point cloud is shown in Fig. 1.

The main advantage of the k-d tree scheme over octree is that it can represents a hierarchical block structure with approximate the same number of points that lends itself to the subsequent transform coding pipeline. Once the k-d tree depth $d$ is determined, the number of points in transform blocks are determined. K-d tree avoids creating empty blocks and sub-graphs, which is significate for further graph transform.

### B. Laplacian Sparsity Optimization for Graph Transform

Graphs are natural representations of 3D irregular point clouds. Comparing with the JPEG grid in [5], graphs preserve more underlying information about the real 3D structure and the correlations among points. Compared with DCT, graph transform is a more data-adaptive method for reducing the spatial redundancy of attribute information.

After k-d tree partition in the point cloud, we form a graph by connecting all points with edges for each transform block. A simple graph is formed as in Fig. 2 (a). Define the graph as $\mathcal{G} = (\mathcal{V} = \{ n_1, n_2, n_3, n_4, n_5 \}, \varepsilon)$, $n_i$ represents the node in the graph $\mathcal{G}$ and $\varepsilon$ represents the sets of edges. The edge weight between two nodes $n_i$ and $n_j$ is:

$$\omega_{i,j} = \begin{cases} e^{-\frac{||n_i - n_j||_2^2}{\sigma^2}}, & if \ ||n_i - n_j||_2^2 \leq \tau; \\ 0, & else, \end{cases} \quad (1)$$

where $\sigma$ denotes the variance of graph nodes and $\tau$ is the Euclidean distance threshold between two nodes. The adjacency matrix $W$ describing the node edge weights for the graph is represented in Fig. 2(b). This pair of parameters $(\sigma, \tau)$ affects the Laplacian sparsity on the matrix $W$. The degree matrix reflecting the correlation density around points is defined as a diagonal matrix $S = diag(s_1, ..., s_5)$, whose element $s_i$ is the sum of elements in $i$th row of $W$.

We choose the graph Laplacian matrix $L$ as the graph shift operator and get the eigen-decomposition of $L$:

$$L = S - W, \quad (2)$$

$$L = A \Lambda A^{-1}, \quad (3)$$

where $A$ is the eigenvector matrix used as the transform matrix and $\Lambda$ is a diagonal matrix including eigenvalues of $L$. The performance of graph transform is closely relevant with the Laplacian matrix $L$, which is associated with the parameters $\sigma$ and $\tau$ in Equation (1). However, current work usually adopt fixed values to set $(\sigma, \tau)$ [7] [12].

To get better performance, we proposed two methods based on online training and offline training respectively to optimize $\sigma$ and $\tau$. For the better training efficiency, we limit the value range of parameter to (0, 1). We set the following variable $f$ and $t$ to represent $\sigma$ and $\tau$:

$$f = e^{-\frac{\bar{d}(n_i, n_j)}{\sigma^2}}, \quad (4)$$

$$t = e^{-\frac{d(n_i n_j)}{\sigma^2}}, \quad (5)$$

where $\bar{d}(n_i, n_j)$ means average distance between two points in the block and $d(n_i n_j)$ means the distance between $n_i$ and $n_j$.

The two training methods are one-pass processing. For online training, we select randomly several transform blocks from the point cloud and traverse the value range to find optimum $f_o$ and $t_o$, which make the performance of graph transform reach the best. Afterward, $f_o$ and $t_o$ are adopted for the following transform blocks. In offline training processing, some typical point clouds except the one to be processed are chosen to be training datasets. We get empirical values $f_o$ and $t_o$ on those datasets and adopt them to process current point cloud.

Online training may get better compression performance than offline training, however, it is impractical to apply it in real-time applications, since it needs to pass all parameters to the decoder and has longer time delay. Instead, offline training has comparable performance without overhead for passing parameters. Therefore, we choose offline training method.

### C. Rate-Distortion Optimization with Lagrangian Method

Laplacian optimized graph transform gives us the signal adaptive energy compaction transform that presents residual coefficients for the quantization and entropy coding. We use different quantization mode by preserving different dimensions and zeroing out the others in the residual matrix, thus result in different R-D performance. To solve the trade-off problem between bitrate and distortion, to obtain the convex-hull optimal quantization mode, we apply the standard Lagrangian method, which is widely accepted in video coding RDO.

For example, a point cloud contains $M$ transform blocks and each transform blocks includes $N$ points, that is, each graph in a block has $N$ nodes and the size of graph transform matrix $A$ is $N \times N$. We define Y component for the point cloud as the matrix $Y$ with $M \times N$ dimensions. The residual matrix $C$ after graph transform is:

$$C_{M \times N} = Y_{M \times N} * A_{N \times N}. \quad (6)$$

Most coefficients in the residual matrix are small, especially on the high dimensions. For lossy point cloud compression, neglecting a part of high-frequency coefficients is a reasonable and efficient strategy. Keeping the first $x$ dimensions in residual matrix $C$ and zeroing out the other dimensions, we can get quantized residual matrix $B(x)$, defined as:

$$b_{i,j} = \begin{cases} c_{i,j}, & if\ j \leq x; \\ 0, & else. \end{cases} \quad (7)$$

Different choice of $x$ means different quantization modes, which may lead to different bitrate $R(x)$ and distortion $D(x)$. Our goal is finding a suitable dimension $x_s$ to get better R-D performance, which is expressed as:

$$x_s = \arg\ min\ D(x),\ s.t.\ R(x) \leq R_{max}, \quad (8)$$

where $R(x)$ is the bitrate after the quantization and $D(x)$ is the sum of absolute difference (SAD) or the sum of square difference (SSD).

A Lagrangian multiplier $\lambda$ is introduced to relax the constraints in Equation (8) and then obtain the following Lagrangian function[13]:

$$L(x) = D(x) + \lambda \times R(x). \quad (9)$$

The goal is finding optimum $x_o$ to make the value of $L(x)$ minimum. Refer to the RDO schemes in traditional video coding [13], the Lagrangian multiplier $\lambda$ has the following relation with the quantization parameter (QP):

$$\lambda = m \times 2^{QP/6}, \quad (10)$$

TABLE I
COMPARISION OF R-D PERFORMANCE ON FOUR DATASETS

| Testing Datasets | Proposed Method | | PCC(DCT-based) | |
|---|---|---|---|---|
| | Y-PSNR (dB) | Bitrate (bpp) | Y-PSNR (dB) | Bitrate (bpp) |
| **1.** *longdress_vox10_1051* | 33.65 | 0.89 | 32.56 | 2.76 |
| **2.** *loot_vox10_1000* | 35.18 | 0.29 | 37.81 | 1.41 |
| **3.** *redandblack_vox10_1451* | 34.38 | 0.55 | 36.1 | 1.54 |
| **4.** *soldier_vox10_0537* | 36.02 | 0.46 | 35.71 | 1.99 |

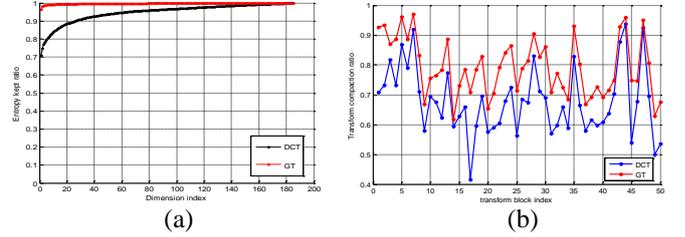

(a)      (b)

Fig. 3 (a) Comparison of transform residual variance between proposed method and DCT on testing 1. It implies the entropy compaction comparison after the transform. (b) Comparison of transform efficiency between proposed method and DCT with 20 dimensions kept in 50 transform blocks of testing 1.

where $m$ is a constant need to be trained by experiments. The Lagrangian multiplier $\lambda$ is only affected by QP, that is, once QP is determined, the optimal quantization mode with the lowest $L$ would be selected.

### III. EXPERIMENTAL RESULTS

We have conducted many tests using frames extracted from 8i dynamic point cloud datasets [14]. We used frames "*longdress_vox10_1300.ply*", "*redandblack_vox10_1550.ply*" "*loot_vox10_1200.ply*", and "*soldier_vox10_0690.ply*" as the training datasets, for which the number of points is 857966, 757691, 805285 and 1089091, and the k-d tree depth $d$ is 13, 12, 12 and 13, respectively. We used four testing frames shown in Table 1. We converted the RGB attributes to $YC_bC_r$ color space and used the luminance Y as the attribute to be coded.

The number of points in each transform block was limited to the empirical range (100, 200), considering the compression performance and the computation complexity. We acquired the optimum parameters $f_o = 0.3$ and $t_o = 0.6$ through the offline training. To select a better quantization mode, we tried $m$ in Equation (10) with some values commonly used in traditional video coding [15] and finally get the trained $m=0.85$ for Lagrangian optimization. Then we used a simple arithmetic encoder by assuming the residual coefficients with a zero-mean Laplacian probability distribution [6]. After the training, we adopted these trained parameters for testing.

The testing R-D performance comparison with Rufael's PCC software [5] based on DCT is shown in Table 1. For all tests, our proposed scheme outperforms DCT. For example, for a luminance peak signal-to-noise ratio (PSNR) on testing 4 around 36dB, DCT based method need around 1.99 bits per point (bpp), while our encoder requires less than 0.46 bpp. That is, the coded bitstream by DCT based encoder is more than 4 times larger than that coded by the proposed encoder.

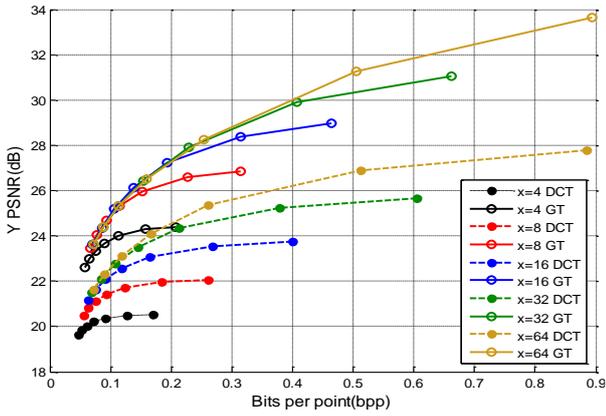

Fig. 4 Compression performance (in dB) vs. bitrate (in bpp) on testing 1, using the proposed encoder and DCT based encoder with five quantization mode at QP=8, 16, 32, 48, 64, 80, 96.

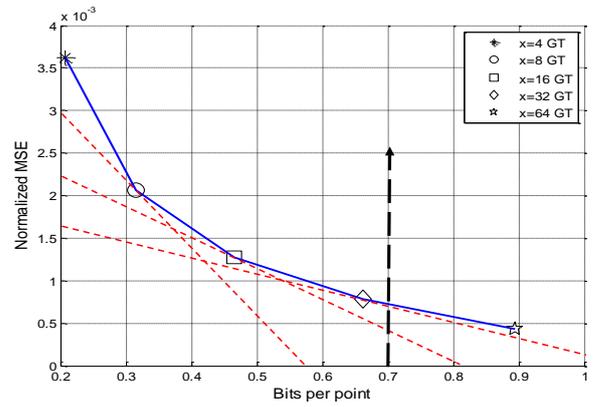

Fig. 5 Lagrangian optimization at QP=8 on testing 1 when $R_{max}$ is given as 0.7 bpp.

For testing 1 with 765821 points, we set $d$=12. There are 4096 transform blocks and each block has 186 or 187 points. Comparison of entropy compaction and transform efficiency between optimized graph transform and DCT is shown in Fig. 3. As we know, the differential entropy for a Laplacian probability distribution variable is directly associated with the Laplacian scale parameter $\Delta$. In video coding, $\Delta$ has an underlying relationship with the variance of transform residuals. Based on this, Fig. 3(a) represents the comparison of entropy compaction efficiency. It tells us that, as keeping same dimensions for graph transform and DCT, the former preserves more information entropy than DCT. When the transform residuals are ranked in a decreasing order of their absolute values [15], calculating the ratio of the sum of several the largest coefficients in the sum of all coefficients is another method to evaluate transform efficiency. We kept the first 20 dimensions in 50 transform blocks and the results are shown in Fig. 3(b). It presents that the optimized graph transform has better transform efficiency over DCT in the blocks.

R-D performance for testing 1 is shown in Fig. 4. Five different quantization modes (x=4,8,16,32,64) with seven QP for uniform quantization are applied. From the results, we can see that optimized graph transform significantly outperforms DCT at all quantization modes and the PSNR difference can be up to 1-3 dB at the same bitrate. The Lagrangian optimization at QP=8 is presented in Fig. 5. It illustrates the processing of operating points convex-hull optimization when the maximum bitrate is given. The quantization mode x=32 is determined as the best mode, which is consistent with the results in Fig. 4.

## IV. CONCLUSIONS

In this paper, we presented an optimized graph transform based scheme for point cloud attribute compression. We optimize the partition for transform blocks to avoid sub-graphs with k-d tree, optimize the Laplacian sparsity for graph transform performance using offline training and optimize the quantization mode selection by Lagrangian RDO. The experimental resultsshowed a significate improvement on transform efficiency and R-D performance. In future work, we plan to do attribute intra and inter prediction based on k-d tree and use our approach for dynamic 3D point cloud sequences.

REFERENCES

[1] J. Kammerl, N. Blodow, R. B. Rusu, S. Gedikli, M. Beetz, and E. Steinbach, "Real-time compression of point cloud streams," in *Proc. IEEE Int. Conf. Robot. Autom.*, May 2012, pp. 778–785.
[2] J. Peng, Chang-Su Kim, and C. C. Jay Kuo, "Technologies for 3d mesh compression: A survey," *Journal of Vis. Comun. and Image Represent.*, vol. 16, no. 6, pp. 688–733, December 2005.
[3] R. Schnabel and R. Klein, "Octree-based point-cloud compression," in *Symposium on Point-Based Graphics*, July 2006.
[4] Y. Huang, J. Peng, C.-C. J. Kuo, and M. Gopi, "A generic scheme for progressive point cloud coding," *IEEE Trans. Vis. Comput. Graphics*, vol. 14, no. 2, pp. 440–453, Mar. Apri. 2008.
[5] R. N. Mekuria, K. Blom, and P. Cesar, "Design, Implementation and Evaluation of a Point Cloud Codec for Tele-Immersive Video," *IEEE Trans. CSVT*, vol. PP, no. 99, pp. 1–1, 2016.
[6] C. Zhang, D. Florêncio, and C. Loops, "Point cloud attribute compression with graph transform"," in *Proc. IEEE Int. Conference on Image Processing*, Paris, France, Sept 2014.
[7] R. A. Cohen, D. Tian, and A. Vetro, "Attribute compression for sparse point clouds using graph transforms," in *2016 IEEE International Conference on Image Processing (ICIP)*, Sept 2016, pp. 1374–1378.
[8] R. L. de Queiroz and P. A. Chou, "Compression of 3d point clouds using a region-adaptive hierarchical transform," *IEEE Transactions on Image Processing*, vol. 25, no. 8, pp. 3947–3956, Aug 2016.
[9] R.Mekuria, P. Cesar, "MP3DG-PCC, Open Source Software Framework for Implementation and Evaluation of Point Cloud Compression," document MPEG 2016/m36527, ISO/IECJTC1/SC29/WG11 Geneva, June 2016.
[10] J. Bentley, "Multidimensional binary search trees used for associative searching," *Communications of the ACM*, 18(9):509–517, Sept 1975.
[11] K. Alsabti, S. Ranka, and V. Singh, "An efficient k-means clustering algorithm," *In 11th Intl. Parallel Processing Symposium*, Mar.1998.
[12] S. Chen, D. Tian, C. Feng, A. Vetro and J. Kovačević, "Fast Resampling of 3D Point Clouds via Graphs," arXiv preprint arXiv:1702.06397v1.
[13] T. Wiegand, H. Schwarz, A. Joch, F. Kossentini, and G. J. Sullivan, "Rate-constrained coder control and comparison of video coding standards," *IEEE Trans. CSVT*, vol. 13, pp. 688–703, July 2003.
[14] E. d'Eon, B. Harrison, T. Myers, P. A. Chou, "8i Voxelized Full Bodies, version 2–A Voxelized Point Cloud Dataset," document MPEG 2017/m74006, ISO/IECJTC1/SC29/WG11 Geneva, January 2017.
[15] Y. Yuan, I.-K. Kim, X. Zheng, L. Liu, and X. Cao, "Quadtree based non-square block structure for inter frame coding in HEVC," *IEEE Trans. CSVT*, vol. 22, no. 12, pp.1707–1719, Dec. 2012.